\documentclass[amsmath,amssymb,aps,prx,reprint,superscriptaddress]{revtex4-2}
\usepackage{amsmath}
\usepackage[table]{xcolor}
\usepackage{amsfonts,amssymb}
\usepackage{graphicx}
\usepackage{hyperref}
\usepackage{academicons}
\usepackage{multirow}
\usepackage[version=4]{mhchem}
\usepackage{makecell}
\usepackage{pgf}
\usepackage{booktabs}
\usepackage[english]{babel}
\usepackage[normalem]{ulem}
\usepackage{natbib}
\usepackage{bm}

\usepackage[capitalise,nameinlink]{cleveref}
\crefname{figure}{Fig.}{Figs.}
\Crefname{figure}{Fig.}{Figs.}
\crefname{equation}{Eq.}{Eqs.}
\Crefname{equation}{Eq.}{Eqs.}
\creflabelformat{equation}{(#2#1#3)}
\crefname{table}{Table}{Tables}
\Crefname{table}{Table}{Tables}
\crefname{edfigure}{Extended~Data~Fig.}{Extended~Data~Figs.}
\Crefname{edfigure}{Extended~Data~Fig.}{Extended~Data~Figs.}

\newcommand{\CellMin}{0}
\newcommand{\CellMax}{0.4}
\colorlet{cmin}{green!10!white}  
\colorlet{cmed}{yellow!20!white} 
\colorlet{cmax}{red!20!white}    
\newcommand{\ColorCell}[1]{%
  \begingroup
  \pgfmathsetmacro{\val}{#1}%
  \pgfmathsetmacro{\norm}{(\val-\CellMin)/(\CellMax-\CellMin)}%
  \pgfmathsetmacro{\norm}{max(min(\norm,1),0)}%
  \pgfmathsetmacro{\pct}{\norm*100}%
  \pgfmathtruncatemacro{\pctint}{round(\pct)}%

  \ifnum\pctint<50
    \pgfmathtruncatemacro{\mix}{round(100 - 2*\pctint)}%
    \edef\temp{\noexpand\cellcolor{cmin!\mix!cmed}}%
  \else
    \pgfmathtruncatemacro{\mix}{round(200 - 2*\pctint)}%
    \edef\temp{\noexpand\cellcolor{cmed!\mix!cmax}}%
  \fi
  \temp #1%
  \endgroup
}

\definecolor{SolidHue}{HTML}{F06292}   
\definecolor{LiquidHue}{HTML}{FDBA74}  
\newcommand{\SolidMin}{58}
\newcommand{\SolidMax}{63}
\newcommand{\LiquidMin}{70}
\newcommand{\LiquidMax}{75}
\newcommand{\DiffMin}{6.4}
\newcommand{\DiffMax}{7.6}
\newcommand{\SolidCell}[1]{%
  \begingroup
  \pgfmathsetmacro{\hnorm}{max(min((#1-\SolidMin)/(\SolidMax-\SolidMin),1),0)}%
  \pgfmathtruncatemacro{\hpct}{round(\hnorm*100)}%
  \edef\temp{\noexpand\cellcolor{SolidHue!\hpct!white}}%
  \temp #1%
  \endgroup
}
\newcommand{\LiquidCell}[1]{%
  \begingroup
  \pgfmathsetmacro{\hnorm}{max(min((#1-\LiquidMin)/(\LiquidMax-\LiquidMin),1),0)}%
  \pgfmathtruncatemacro{\hpct}{round(\hnorm*100)}%
  \edef\temp{\noexpand\cellcolor{LiquidHue!\hpct!white}}%
  \temp #1%
  \endgroup
}
\newcommand{\DiffCell}[1]{%
  \begingroup
  \pgfmathsetmacro{\hnorm}{max(min((#1-\DiffMin)/(\DiffMax-\DiffMin),1),0)}%
  \pgfmathtruncatemacro{\hpct}{round(\hnorm*100)}%
  \edef\temp{\noexpand\cellcolor{SolidHue!\hpct!white}}%
  \temp #1%
  \endgroup
}
\newcommand{\DiffCellV}[2]{%
  \begingroup
  \pgfmathsetmacro{\hnorm}{max(min((#1-\DiffMin)/(\DiffMax-\DiffMin),1),0)}%
  \pgfmathtruncatemacro{\hpct}{round(\hnorm*100)}%
  \edef\temp{\noexpand\cellcolor{SolidHue!\hpct!white}}%
  \temp #2%
  \endgroup
}

\begin{document}

\title{Linking information entropy to atomistic thermodynamics by dynamically coarse-graining graph latent representations}

\author{Mengyi Chen}
\thanks{These authors contributed equally.}
\affiliation{Department of Materials Science and Engineering, National University of Singapore, Singapore}
\affiliation{Department of Mathematics, National University of Singapore, Singapore}

\author{Xintian Wang}
\thanks{These authors contributed equally.}
\affiliation{Department of Materials Science and Engineering, National University of Singapore, Singapore}

\author{Zihan Zhang}
\thanks{These authors contributed equally.}
\affiliation{Department of Materials Science and Engineering, National University of Singapore, Singapore}

\author{Qianxiao Li}
\affiliation{Department of Mathematics, National University of Singapore, Singapore}
\affiliation{Institute for Functional Intelligent Materials, National University of Singapore, Singapore}

\author{Peichen Zhong}
\email[]{zhongpc@nus.edu.sg}
\affiliation{Department of Materials Science and Engineering, National University of Singapore, Singapore}
\affiliation{Institute for Functional Intelligent Materials, National University of Singapore, Singapore}

\date{\today}

\begin{abstract}
Predicting phase stability, transition temperatures, and polymorphic transformations requires accurate entropy and free-energy estimation.
The thermodynamic entropy separates by timescale into electronic, vibrational and configurational parts, and the configurational part, set by slow hops between metastable configurations, remains the main computational bottleneck.
Information entropy, by contrast, is easy to compute for any probability distribution over a set of states, but acquires a thermodynamic meaning only when the states are separated by slow hops.
Here we introduce EntroVAMP, a site-free approach for computing entropy that learns coarse-grained atomic states from the slow motions in a molecular dynamics trajectory.
A graph neural network encodes each atomic environment and is trained with the variational approach for Markov processes to find the metastable states that persist over time.
The configurational entropy is then computed as the information entropy of these states, conditioned on the states of neighboring atoms.
EntroVAMP reproduces the calorimetric transition entropy of the superionic conductor AgI and, combined with thermodynamic integration, recovers the polymorph stability sequence and transition regimes of the solid electrolyte \ce{Li3PS4}.
Beyond thermodynamics, the information entropy serves as a collective variable for enhanced sampling, driving superionic AgI into its hexagonal, cubic and stacking-disordered forms.
By linking information entropy to thermodynamics, EntroVAMP makes material entropy computable directly from molecular dynamics, bringing finite-temperature free energies and rare-event phase transitions within reach of atomistic simulation.
\end{abstract}


\maketitle
\section{Introduction}

\begin{figure*}[!t]
\centering
\includegraphics[width=\textwidth]{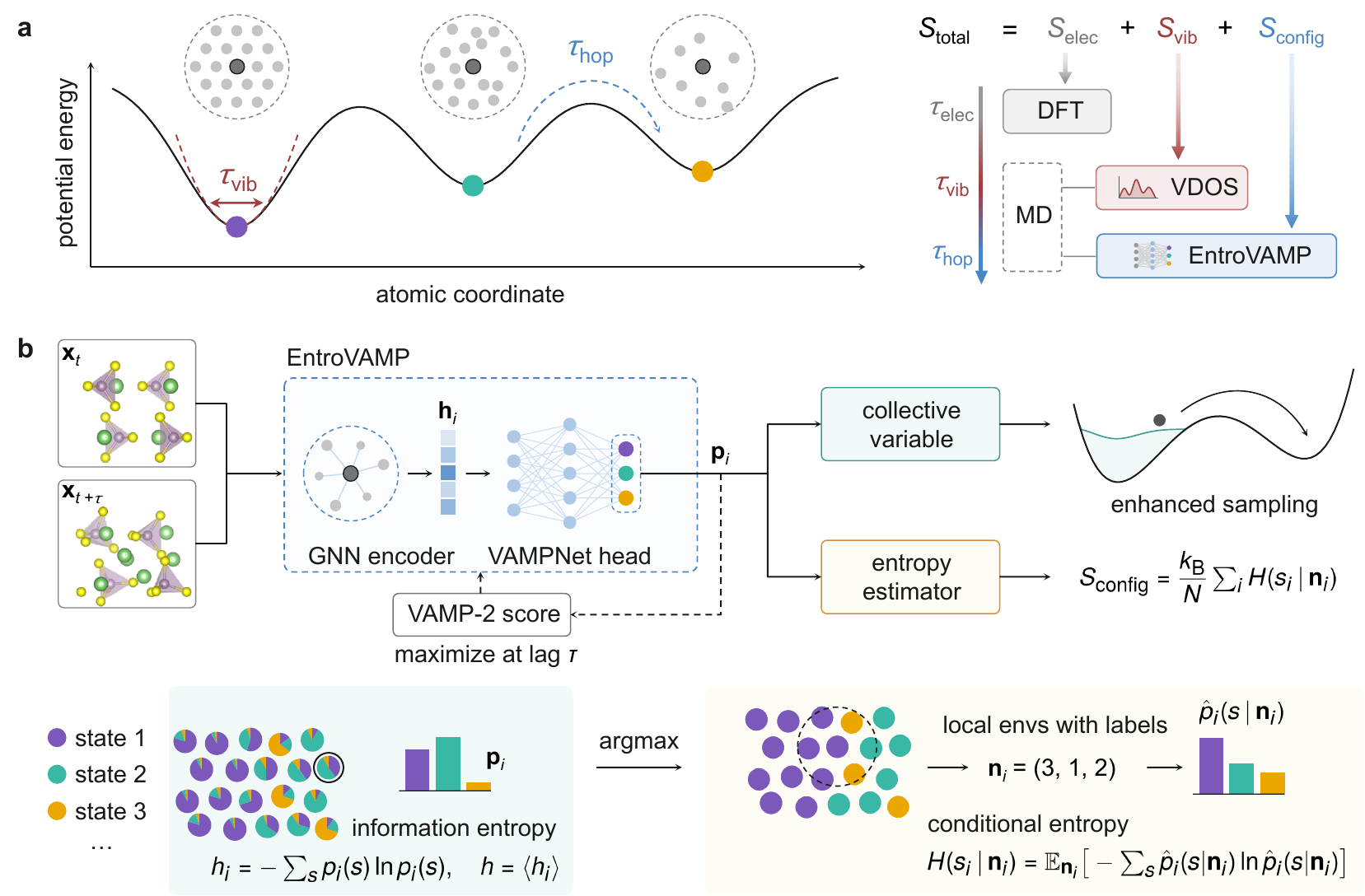}
\caption{\textbf{Entropy decomposition by timescale and the EntroVAMP framework.}
\textbf{a}, Schematic potential-energy surface (PES) along an atomic coordinate. Each basin corresponds to a metastable local environment (insets, central atom in dark gray). Electronic excitations occur on the shortest timescale ($\tau_{\mathrm{elec}}$), intra-basin vibrations on an intermediate timescale ($\tau_{\mathrm{vib}}$), and inter-basin hops on a much longer timescale ($\tau_{\mathrm{hop}}$). This timescale separation decomposes the total entropy as $S = S_{\mathrm{elec}} + S_{\mathrm{vib}} + S_{\mathrm{config}}$, with $S_{\mathrm{elec}}$ obtained from density functional theory (DFT), $S_{\mathrm{vib}}$ from the vibrational density of states (VDOS) of molecular dynamics (MD) trajectories, and $S_{\mathrm{config}}$ from EntroVAMP applied to the same trajectories.
\textbf{b}, In EntroVAMP, an equivariant graph neural network (GNN) encodes structures $\mathbf{x}_t$ and $\mathbf{x}_{t+\tau}$, separated by a lag time $\tau$, into local-environment embeddings $\mathbf{h}_i$ for each atom $i$. Species-specific VAMPNet heads map $\mathbf{h}_i$ to soft state probabilities $\mathbf{p}_i$ by maximizing the VAMP-2 score at lag $\tau$. From $\mathbf{p}_i$, EntroVAMP constructs both collective variables (CVs) for enhanced sampling and an estimator of $S_{\mathrm{config}}$. 
Lower left: $\mathbf{p}_i$ is represented as a pie chart per atom (state colors in the key); $h_i$ is the per-atom information entropy and $h = \langle h_i \rangle$ is its species average. 
Lower right: each atom is assigned its most probable state $s_i = \arg\max_s p_i(s)$, the states of its first-shell same-species neighbors (dashed circle) are tallied in the count vector $\mathbf{n}_i$, and $S_{\mathrm{config}}$ is evaluated from the conditional Shannon entropy $H(s_i \mid \mathbf{n}_i)$ of the empirical distribution $\hat{p}_i(s \mid \mathbf{n}_i)$.}
\label{fig:pipeline}
\end{figure*}

Free energy governs the thermodynamic behavior of matter across materials science, chemistry, and biology, determining phase stability, transition temperatures, latent heats, and the free-energy barriers that control transition kinetics \cite{cheng_ab_2019, aznar2017giant, sosso_crystal_2016}.
In the free energy $F = U - TS$, the internal energy $U$ favors low-energy ordered structures, whereas the entropic term $-TS$ (where $T$ is the temperature and $S$ is the entropy) increasingly favors disordered states as temperature rises.
In molecular dynamics (MD) simulations, $U$ can be computed directly as a time average of the frame energies, whereas entropy is a collective property of the probability distribution over high-dimensional phase space and cannot be expressed as the average of a single-frame observable.
High-throughput materials discovery pipelines therefore typically rank candidates by their ground-state ($0$~K) energies relative to the convex hull \cite{merchant_scaling_2023, riebesell_framework_2025}, a limit in which the entropic term vanishes.
However, predicting polymorph stability and guiding synthesis require free energies at elevated temperatures~\cite{park_closing_2026}, where entropy stabilizes high-temperature polymorphs~\cite{kam_crystal_2023}, superionic conductors~\cite{boyce_huberman_1979}, and high-entropy ceramics~\cite{oses_high-entropy_2020}.

In condensed phases, the total entropy is commonly decomposed by the timescales of the underlying degrees of freedom \cite{ceder_derivation_1993}:
\begin{equation}
    S = S_{\mathrm{elec}} + S_{\mathrm{vib}} + S_{\mathrm{config}}.
    \label{eq:decomp}
\end{equation}
The electronic contribution $S_{\mathrm{elec}}$ is usually small, except in metals at high temperatures \cite{zhang_accurate_2017} and in mixed-valence compounds where localized charges disorder across lattice sites \cite{zhou_configurational_2006}.
Separating $S_{\mathrm{vib}}$ from $S_{\mathrm{config}}$ relies on partitioning microstates into coarse-grained discrete configurations $\boldsymbol{\sigma}$, i.e., basins on the potential-energy surface within which atoms vibrate on a fast timescale $\tau_{\mathrm{vib}}$ and between which the system transitions only by slow hops on a timescale $\tau_{\mathrm{hop}} \gg \tau_{\mathrm{vib}}$ (Fig.~\ref{fig:pipeline}a).
The vibrational term $S_{\mathrm{vib}}$, which measures the entropy of microstates explored within a single configurational basin, is routinely evaluated from lattice-dynamical phonon calculations or from the vibrational density of states (VDOS) of MD trajectories.
The configurational term, $S_{\mathrm{config}} = -k_B \sum_{\boldsymbol{\sigma}} p(\boldsymbol{\sigma}) \ln p(\boldsymbol{\sigma})$, is the Gibbs--Shannon entropy of the probability distribution $p(\boldsymbol{\sigma})$ over the discrete configurations of one phase, where $k_B$ is the Boltzmann constant.
While $S_{\mathrm{config}}$ vanishes for a defect-free ordered crystal, it often dominates the entropy difference between competing phases in substitutionally disordered solids, superionic conductors, and liquids.
Yet $S_{\mathrm{config}}$ remains the hardest term in Eq.~\eqref{eq:decomp} to evaluate, because its value depends on how continuous atomic coordinates are partitioned into discrete configurations.

Existing approaches to finite-temperature thermodynamics differ in how they define and sample configurational states.
The conventional lattice route maps the energy onto discrete site occupations of a reference crystal structure via a cluster expansion and samples with Monte Carlo simulations \cite{van_de_walle_alloy_2002, angqvist_icet_2019, barroso-luque_cluster_2022, puchala_casm_2025}.
Although well established for substitutional alloys, this approach presupposes a fixed reference lattice and breaks down when atoms undergo large anharmonic excursions or leave their lattice sites, as in liquids, glasses, and superionic conductors.
To avoid a reference lattice, site-free methods estimate information entropy directly from MD trajectories using local structural descriptors \cite{hong_generalized_2025, schwalbe-koda_model-free_2025}.
For instance, \citet{hong_generalized_2025} define discrete atomic states by the nearest-neighbor coordination number, whereas \citet{schwalbe-koda_model-free_2025} evaluate the kernel-density information entropy of continuous atom-centered descriptors.

However, hand-chosen scalar descriptors such as coordination number may not resolve chemically or geometrically distinct environments, continuous descriptor densities depend on the kernel bandwidth and mix vibrational fluctuations with configurational disorder, and summing per-atom entropies overestimates $S_{\mathrm{config}}$ when the states of neighboring atoms are correlated.
Alternatively, thermodynamic integration bypasses the decomposition in Eq.~\eqref{eq:decomp} by integrating along an artificial reversible path to a reference system of known free energy, such as an Einstein crystal \cite{freitas_nonequilibrium_2016, cheng_computing_2018}.

Here we introduce EntroVAMP, a framework that computes entropy from coarse-grained atomic states learned from the dynamics of MD trajectories, consistent with the statistical-mechanical definition of configurations as basins separated by slow hops (Fig.~\ref{fig:pipeline}b).
Graph neural networks trained with the variational approach for Markov processes (VAMP) \cite{mardtVAMPnetsDeepLearning2018, wuVariationalApproachLearning2020} can identify slowly interconverting states in MD trajectories of materials and biomolecules \cite{xie_graph_2019, ghorbani_graphvampnet_2022, liu_graphvampnets_2023}, helping interpret the dynamics.
By maximizing time-lagged correlations over a lag time $\tau$, the network identifies latent states that persist across vibrations, requiring neither a predefined reference lattice nor hand-crafted descriptors.
EntroVAMP turns the statistics of the learned coarse-grained states into a thermodynamic entropy.
The configurational entropy $S_{\mathrm{config}}$ is evaluated from the learned state assignments, conditioned on neighboring-atom states when local correlations persist, and combined with $S_{\mathrm{vib}}$ and $S_{\mathrm{elec}}$ in Eq.~\eqref{eq:decomp} to obtain the total entropy from a single MD trajectory.
The lifetimes of the learned states also indicate when the estimate can be trusted, because they reveal the temperatures at which hops become as fast as vibrations and the decomposition in Eq.~\eqref{eq:decomp} breaks down.
We test the approach against the calorimetric entropy of the superionic transition in AgI and on the polymorphic transitions of \ce{Li3PS4}, where EntroVAMP supplies the low-temperature reference entropy from which free energies are integrated over temperature.
Beyond equilibrium thermodynamics, the Shannon and relative entropies of the soft state probabilities provide differentiable collective variables (CVs) for enhanced sampling.
Using well-tempered metadynamics, we drive superionic $\alpha$-AgI into $\beta$-AgI, $\gamma$-AgI, and stacking-disordered AgI, a kinetically favored polytype observed experimentally upon cooling~\cite{smithStackingDisorderDesign2019} that was not included in the training set.

\section{Results}

\subsection{Information entropy from generalized configurations}
The Gibbs entropy $S = -k_B\sum_s p(s)\ln p(s)$ can be decomposed as
\begin{equation}
    S = -k_B \sum_{\boldsymbol{\sigma}} p(\boldsymbol{\sigma})\ln p(\boldsymbol{\sigma})
      + \sum_{\boldsymbol{\sigma}} p(\boldsymbol{\sigma})\, k_B H(\cdot\mid\boldsymbol{\sigma}),
    \label{eq:S_split}
\end{equation}
by grouping the microscopic states $s$ into configurations $\boldsymbol{\sigma}$, where $H(\cdot\mid\boldsymbol{\sigma}) = -\sum_{s\in\boldsymbol{\sigma}} p(s\mid\boldsymbol{\sigma})\ln p(s\mid\boldsymbol{\sigma})$ is the Shannon (or information) entropy of the microscopic states within $\boldsymbol{\sigma}$, with $p(s\mid\boldsymbol{\sigma})$ the probability $p(s)$ renormalized over $\boldsymbol{\sigma}$.
We also refer to the Shannon entropy as the information entropy.
Summing the Boltzmann probabilities $p(s) = e^{-\beta E_s}/Z$ within a configuration gives \cite{ceder_derivation_1993}
\begin{equation}
\begin{aligned}
    p(\boldsymbol{\sigma}) &= \sum_{s\in\boldsymbol{\sigma}} p(s) = \frac{e^{-\beta F(\boldsymbol{\sigma})}}{Z},\\
    F(\boldsymbol{\sigma}) &= -\frac{1}{\beta}\ln\sum_{s\in\boldsymbol{\sigma}} e^{-\beta E_s},
\end{aligned}
    \label{eq:config_freeE}
\end{equation}
with $\beta = 1/k_BT$ and $Z = \sum_{\boldsymbol{\sigma}} e^{-\beta F(\boldsymbol{\sigma})}$.
The weight of a configuration is therefore set by a free energy $F(\boldsymbol{\sigma})$ rather than an energy, since the sum over its microscopic states includes its vibrational and electronic excitations.
The thermodynamic $S_{\mathrm{config}}$ is defined by partitioning the system into basins it leaves by slow hops. For this partition, the two terms of Eq.~\eqref{eq:S_split} are the thermodynamic $S_{\mathrm{config}}$ and the entropy within a basin, which Eq.~\eqref{eq:decomp} treats as $S_{\mathrm{vib}} + S_{\mathrm{elec}}$.
Since $S_{\mathrm{vib}}$ can be estimated from the VDOS and $S_{\mathrm{elec}}$ is usually small, the remaining task is to determine $S_{\mathrm{config}}$, particularly when establishing the absolute entropy of a reference state.
The difficulty lies in identifying the configurations $\boldsymbol{\sigma}$ in practice and in sampling $p(\boldsymbol{\sigma})$.

The lattice route presupposes a reference lattice and takes $\boldsymbol{\sigma}$ to be its site occupations \cite{van_de_walle_alloy_2002, angqvist_icet_2019, barroso-luque_cluster_2022, puchala_casm_2025}. 
Because evaluating $F(\boldsymbol{\sigma})$ for every configuration is computationally prohibitive, conventional lattice models replace $F(\boldsymbol{\sigma})$ with the static energy $E(\boldsymbol{\sigma}_0)$ of the nearest relaxed structure in Monte Carlo simulations to sample $p(\boldsymbol{\sigma})$. 
While this approximation often reproduces the qualitative topology of alloy phase diagrams, neglecting intra-basin vibrational excitations and off-lattice anharmonic excursions shifts order--disorder transition temperatures and destabilizes entropically driven phases \cite{garbulsky_effect_1994}.
In contrast, an equilibrium MD trajectory spends a fraction of time in each configuration $\boldsymbol{\sigma}$ proportional to $\sum_{s \in \boldsymbol{\sigma}} e^{-\beta E_s} = e^{-\beta F(\boldsymbol{\sigma})}$, thereby sampling $p(\boldsymbol{\sigma})$ with the full configurational free energy $F(\boldsymbol{\sigma})$ built in.
To evaluate $S_{\mathrm{config}}$ directly from MD, each atom $i$ can be assigned a discrete state $s_i$ that describes its local environment, such as its number of same-element neighbors \cite{hong_generalized_2025}, so that $\boldsymbol{\sigma}$ is described by the states of all atoms.
$S_{\mathrm{config}}$ can then be estimated by the mean per-atom Shannon entropy:
\begin{equation}
    S_{\mathrm{atom}} = \frac{k_B}{N}\sum_{i=1}^{N} H(s_i),
    \qquad H(s_i) = -\sum_{s}\hat{p}_i(s)\ln\hat{p}_i(s),
    \label{eq:satom}
\end{equation}
where $\hat{p}_i(s)$ is the fraction of frames in which atom $i$ is in state $s$ and $N$ the number of atoms.
However, for local states that resolve the slow hops between configurations, Eq.~\eqref{eq:satom} generally overestimates $S_{\mathrm{config}}$ in correlated systems, since it treats the states of different atoms as independent.

\begin{figure*}[t]
\centering
\includegraphics{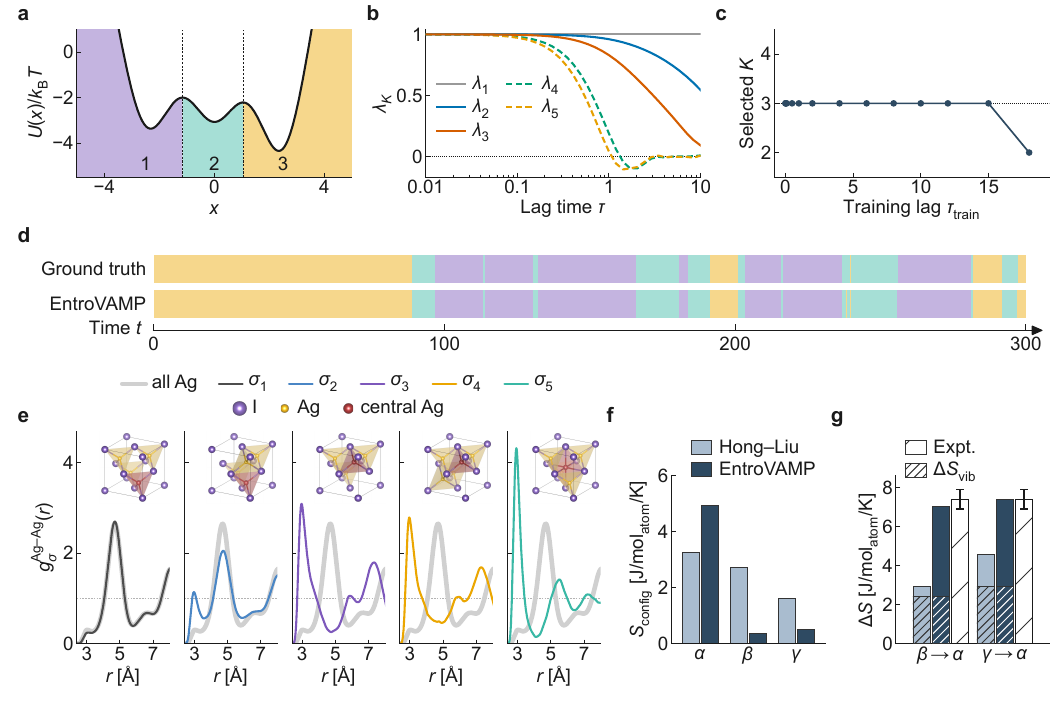}
\caption{
\textbf{EntroVAMP state identification in a three-well potential and entropy estimation for AgI.}
\textbf{a}, Three-well potential $U(x)/k_B T$ and its three ground-truth states (shaded regions 1--3).
\textbf{b}, Transition-matrix eigenvalues $\lambda_K$ versus lag time $\tau$ as the number of states $K$ is increased from 1 to 5. Each increment in $K$ introduces one additional eigenvalue $\lambda_K$, with $\lambda_4$ and $\lambda_5$ (dashed) falling below zero.
\textbf{c}, Number of states $K$ selected by the eigenvalue criterion versus training lag $\tau_{\mathrm{train}}$ (identical across three random seeds).
\textbf{d}, Ground-truth and EntroVAMP ($K=3$) state assignments along a segment of the trajectory, colored as in \textbf{a}.
\textbf{e}, State-resolved Ag--Ag radial distribution functions $g_\sigma^{\mathrm{Ag\text{--}Ag}}(r)$ for the five learned Ag states ($\sigma_1$--$\sigma_5$) of $\gamma$-AgI at 400~K, compared with the overall Ag--Ag distribution (gray). Insets show a representative configuration of each state in the zincblende cell, assigned a posteriori (central Ag in dark red, other Ag in yellow, and I in purple).
\textbf{f}, Configurational entropy $S_{\mathrm{config}}$ of $\alpha$-, $\beta$-, and $\gamma$-AgI at 400~K from EntroVAMP and from the neighbor-count estimator of \citet{hong_generalized_2025} (Hong--Liu).
\textbf{g}, Total entropy change $\Delta S$ of the $\beta\rightarrow\alpha$ and $\gamma\rightarrow\alpha$ transitions of AgI at 400~K from the two methods, compared with experiment (Expt.; error bars indicate experimental uncertainty)~\cite{aznar2017giant}. Hatched and solid portions denote the vibrational ($\Delta S_{\mathrm{vib}}$) and configurational ($\Delta S_{\mathrm{config}}$) contributions, respectively.}
\label{fig:toy_agi}
\end{figure*}

\subsection{Entropy from dynamically coarse-grained states}
\label{subsec:corse_grained}

VAMPNets identify slow coarse-grained states from a trajectory by maximizing the VAMP-2 score \cite{mardtVAMPnetsDeepLearning2018, wuVariationalApproachLearning2020}.
The VAMP-2 score is the sum of squared singular values of the normalized correlation matrix between state assignments separated by a lag time $\tau$, and maximizing it favors assignments that stay correlated over $\tau$.
Prior works have combined graph neural networks (GNNs) and VAMPNets to model and interpret the dynamics of materials and biomolecules \cite{xie_graph_2019, ghorbani_graphvampnet_2022, liu_graphvampnets_2023}.
EntroVAMP adopts the network architecture and training objective of these works but focuses on computing information entropy from statistics of coarse-grained states.

Specifically, EntroVAMP combines a graph equivariant transformer (GET) encoder that embeds each atomistic local environment \cite{bamboo_2025} with species-specific VAMPNet heads.
Each head maps the local-environment embedding of atom $i$ to soft probabilities $\mathbf{p}_i$ over the $K$ coarse-grained latent states of its species, and $K$ may differ by species.
The metastable states differ between phases and change with temperature, so we train a separate EntroVAMP model for each phase and temperature.
When configurational hops on the potential energy surface (PES) are much slower than vibrations, $\tau_{\mathrm{vib}}\ll\tau_{\mathrm{hop}}$, assignments that distinguish metastable environments stay correlated up to $\tau_{\mathrm{hop}}$, whereas assignments that follow the vibrations decorrelate within $\tau_{\mathrm{vib}}$.
Both timescales can be estimated from the mean squared displacement, which stops growing ballistically after about $\tau_{\mathrm{vib}}$ and resumes linear growth after about $\tau_{\mathrm{hop}}$.
The lag time $\tau$ is chosen well below $\tau_{\mathrm{hop}}$ so that the states remain correlated over the lag.
The learned states can then capture slow transitions between configurations rather than vibrations within them.
The eigenvalues $\lambda_k$ of the transition matrix between these states at lag $\tau$ measure how strongly each slow transition persists over $\tau$. To select $K$, we scan candidate state numbers and keep the largest one for which the next eigenvalue $\lambda_{K+1}$ is either unphysically negative or separated from $\lambda_K$ by a large spectral gap (see Methods, `EntroVAMP implementation').

To illustrate the choice of $K$ and the effect of the lag time $\tau$, we apply EntroVAMP to a particle undergoing Langevin dynamics in a three-well potential, $U(x)=c\,x^4-\sum_{k=1}^{3}D_k\exp[-(x-\mu_k)^2/2s^2]$, the sum of three Gaussian wells of depth $D_k$ centered at $\mu_k$ and a confining term $c\,x^4$ (Fig.~\ref{fig:toy_agi}a).
The potential has three metastable states by construction, and EntroVAMP is trained to discover them from the trajectory alone.
When the number of states is increased from 3 to 4, a negative eigenvalue appears, so the eigenvalue criterion recovers the three metastable states (Fig.~\ref{fig:toy_agi}b).
Applied at different training lags, the criterion gives $K=3$ up to $\tau_\mathrm{train}=15$ and loses one state at $\tau_\mathrm{train}=18$, once the fastest hop decorrelates within the lag (Fig.~\ref{fig:toy_agi}c).
For the particle in the three-well potential, the most probable state agrees with the well it occupies (Fig.~\ref{fig:toy_agi}d).

Once the states are learned, EntroVAMP computes the information entropy from their statistics.
At each frame, the local environment of each atom is assigned to its most probable learned state $s_i$, and counting over the frames gives $\hat{p}_i(s)$ and $S_{\mathrm{atom}}$ of Eq.~\eqref{eq:satom} for each species.
When the atoms exchange freely within the observation window, as on the Ag sublattice of superionic $\alpha$-AgI, $S_{\mathrm{config}} = S_{\mathrm{atom}}$.
When the atoms do not exchange freely on the simulated timescale, a case of broken ergodicity \cite{palmerBrokenErgodicity1982, mauroContinuouslyBrokenErgodicity2007}, as for Li in \ce{Li3PS4} at low temperature, the states of neighboring atoms are correlated and $S_{\mathrm{atom}}$ overestimates $S_{\mathrm{config}}$.
We condition the state of each atom on the states of its first-shell neighbors,
\begin{equation}
    S_\mathrm{loc} = \frac{k_B}{N} \sum_{i=1}^{N} H\!\left(s_i \,\middle|\, \mathbf{n}_i\right) .
  \label{eq:sloc_main}
\end{equation}
Here, $\mathbf{n}_i$ counts the states of the $m$ nearest same-species neighbors of atom $i$, with $m$ chosen to cover the first neighbor shell, and $H(s_i\mid\mathbf{n}_i)$ is the conditional entropy of $s_i$ given $\mathbf{n}_i$.
Because conditioning cannot increase entropy, $S_\mathrm{loc} \le S_\mathrm{atom}$, and the two coincide for $m=0$.
We report $S_\mathrm{config} = S_\mathrm{loc}$ when the distributions $\hat{p}_i(s)$ differ measurably from atom to atom, the signature of atoms that do not exchange freely, and $S_\mathrm{config} = S_\mathrm{atom}$ otherwise.
For a compound, the contributions of the species are weighted by their atomic fractions,  $S_\mathrm{config}=\frac{3}{8}S_\mathrm{config}^{(\mathrm{Li})}+\frac{1}{8}S_\mathrm{config}^{(\mathrm{P})}+\frac{4}{8}S_\mathrm{config}^{(\mathrm{S})}$ for \ce{Li3PS4}.

\subsection{AgI superionic phase transition}

The entropy change across a phase transition sets the cooling and heat-storage capacity of caloric materials \cite{aznar2017giant}, making accurate configurational-entropy estimation essential when large disorder drives the transition.
AgI is a prototypical solid electrolyte that undergoes a first-order transition at about 420~K with 25~K hysteresis from the relatively ordered hexagonal $\beta$ and cubic $\gamma$ phases to the highly conducting superionic $\alpha$ phase, where Ag$^{+}$ ions become dynamically disordered over interstitial sites. This superionic transition is driven by the disordering of the Ag sublattice and is accompanied by a large entropy increase of $63 \pm 4~\mathrm{J/kg/K}$, corresponding to $7.4 \pm 0.5~\mathrm{J/mol_{atom}/K}$ \cite{aznar2017giant}. Below the transition, AgI is a mixture of the $\beta$ and $\gamma$ phases, so we compute the $\beta\rightarrow\alpha$ and $\gamma\rightarrow\alpha$ entropy changes separately with EntroVAMP.

Figure~\ref{fig:toy_agi}g compares the entropy changes from EntroVAMP and from a neighbor-count estimator with experiment. For the latter we follow \citet{hong_generalized_2025}: the state of each atom is its number of same-element neighbors within the first minimum of the pair distribution function.
Both methods use the same vibrational entropy from the VDOS, 2.44 and 2.94~J/mol$_\mathrm{atom}$/K for $\beta\rightarrow\alpha$ and $\gamma\rightarrow\alpha$, which accounts for less than half of the measured value.
EntroVAMP adds a configurational entropy change of 4.60 and 4.46~J/mol$_\mathrm{atom}$/K, giving $\Delta S_{\beta\rightarrow\alpha}=7.04$ and $\Delta S_{\gamma\rightarrow\alpha}=7.40~\mathrm{J/mol_{atom}/K}$, both within the experimental uncertainty.
The neighbor-count estimator adds 0.52 and 1.64~J/mol$_\mathrm{atom}$/K, and its totals of 2.96 and 4.58~J/mol$_\mathrm{atom}$/K fall below the measured value.
The two estimators differ in both directions (Fig.~\ref{fig:toy_agi}f). For superionic $\alpha$, EntroVAMP gives 4.96~J/mol$_\mathrm{atom}$/K and the neighbor-count estimator 3.26~J/mol$_\mathrm{atom}$/K. For the ordered $\beta$ and $\gamma$ phases, EntroVAMP gives 0.36 and 0.50~J/mol$_\mathrm{atom}$/K, whereas the neighbor-count estimator gives 2.74 and 1.62~J/mol$_\mathrm{atom}$/K. 
The small but finite values that EntroVAMP assigns to $\beta$ and $\gamma$ may originate from the low-lying, nearly flat Ag optical modes and large-amplitude Ag rattling \cite{BUHRER1975155, PhysRevB.107.064308}, which may favor excursions of Ag into interstitial sites.

To illustrate the learned states and these excursions of Ag into interstitial sites, we take $\gamma$-AgI as an example (Fig.~\ref{fig:toy_agi}e). We map each state a posteriori onto a representative configuration in the zincblende cell, with the central Ag on its most probable site type. The central Ag occupies a lattice tetrahedral site in $\sigma_1$ and $\sigma_2$, with an interstitial Ag next to it in $\sigma_2$. In $\sigma_3$ and $\sigma_4$ the central Ag occupies an interstitial tetrahedral site, which is empty in the ideal lattice, next to a vacancy, with more interstitial Ag and vacancies nearby in $\sigma_4$. In $\sigma_5$ the central Ag occupies an interstitial octahedral site.
The coarse-grained states learned by EntroVAMP thus differ in the relative positions of the neighboring atoms, both the I cage around the central Ag and the nearby interstitial Ag and vacancies, differences that neighbor numbers cannot capture.
For superionic $\alpha$, EntroVAMP and the neighbor-count estimator both give a large configurational entropy.
In the ordered $\beta$ and $\gamma$ phases, where the Ag atoms rarely leave their sites, the neighbor-count estimator instead gives three to eight times the EntroVAMP value.
The neighbor-count estimator places its cutoff at the first minimum of the Ag--Ag pair distribution function, at 3.95~\AA{} in $\beta$ and 3.35~\AA{} in $\gamma$, between the short Ag--Ag contacts near 3~\AA{} and the first Ag lattice shell near 4.7~\AA{} (SI Sec.~S4 and Fig.~S10).
The neighbor number then counts these short-lived contacts rather than the slow occupation of the Ag sites that EntroVAMP resolves.

\subsection{Entropic effects in \ce{Li3PS4} polymorphs}

\begin{figure*}[t]
\centering
\includegraphics{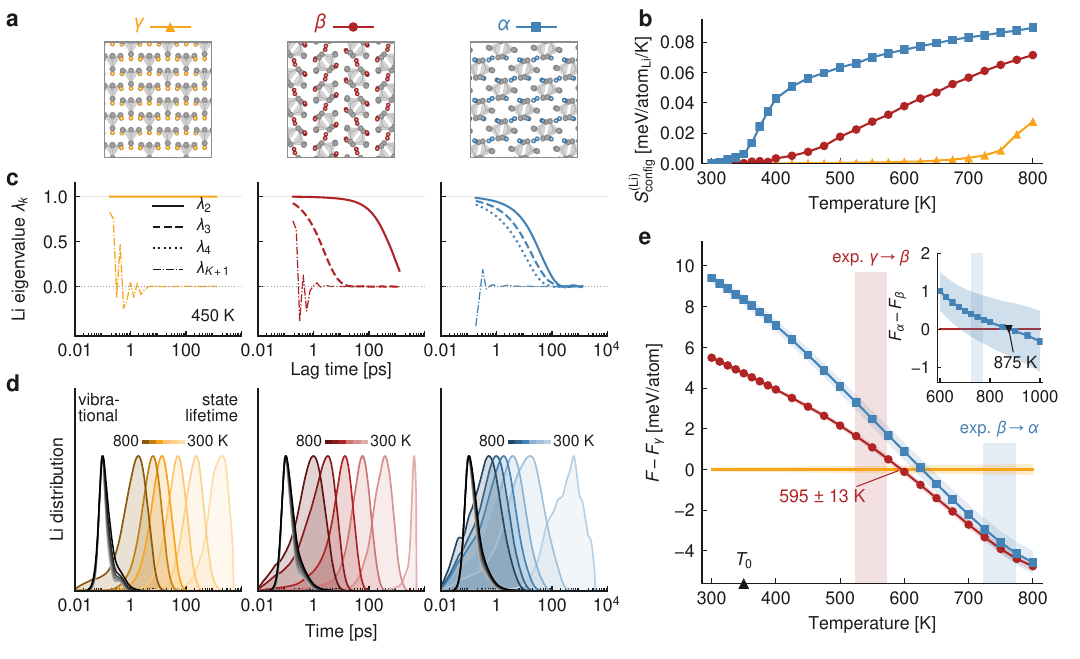}
\caption{\textbf{Entropy and phase stability of the \ce{Li3PS4} polymorphs.}
\textbf{a}, Crystal structures of $\gamma$- (yellow), $\beta$- (red), and $\alpha$-\ce{Li3PS4} (blue), with Li atoms in the color of each polymorph and \ce{PS4} tetrahedra in gray; the same color and marker scheme applies throughout \textbf{b}--\textbf{e}.
\textbf{b}, Configurational entropy of Li from EntroVAMP, $S_\mathrm{config}^{(\mathrm{Li})}$ (meV/atom$_\mathrm{Li}$/K), versus temperature.
\textbf{c}, Eigenvalues $\lambda_k$ of the Li transition matrix versus lag time at 450~K for $\gamma$ ($K=2$, left), $\beta$ ($K=3$, middle), and $\alpha$ ($K=4$, right). The dash-dot curve ($\lambda_{K+1}$) is the additional eigenvalue from a model trained with $K+1$ states.
\textbf{d}, Distributions of the Li vibrational period (gray to black curves) and duration-weighted Li state lifetime (filled curves, light to dark from 300 to 800~K in steps of 100~K) for $\gamma$, $\beta$, and $\alpha$ (left to right), each normalized to a peak height of one.
\textbf{e}, Free energies of $\beta$ and $\alpha$ relative to $\gamma$ ($F - F_\gamma$) integrated from the reference temperature $T_0 = 350$~K, with shaded bands indicating two standard errors; pale vertical bands mark the experimental $\gamma\rightarrow\beta$ (523--573~K) and $\beta\rightarrow\alpha$ (723--773~K) transition ranges. Inset: $F_\alpha-F_\beta$ (meV/atom) up to 1000~K, crossing zero at 875~K (black triangle), with the lower edge of its uncertainty band reaching zero at 680~K. All trajectories were simulated at 1~bar using MACE-MH-1 fine-tuned to r$^2$SCAN+rVV10.}
\label{fig:lps_stability}
\end{figure*}

At room temperature, the solid electrolyte \ce{Li3PS4} adopts the ordered $\gamma$ polymorph, which is a poor Li-ion conductor \cite{gigli_mechanism_2024, miura_time_temperature_2026}. 
Upon heating, $\gamma$-\ce{Li3PS4} transforms into the superionic $\beta$ polymorph at 523--573~K and into the $\alpha$ polymorph at 723--773~K (Fig.~\ref{fig:lps_stability}a) \cite{homma_crystal_2011}. 
Static DFT calculations at the r$^2$SCAN+rVV10 level~\cite{furness_accurate_2020, ning_workhorse_2022} reproduce this ground-state energy ordering, placing $\beta$ and $\alpha$ only 7.9 and 11.8~meV/atom above $\gamma$ (SI Table~S3). \ce{Li3PS4} provides a stringent test of when entropy separates into vibrational and configurational contributions and how reference-entropy governs predicted polymorph stability. 
We performed MD simulations of all three polymorphs using a MACE-MH-1 potential \cite{Batatia2025MACEMH1} fine-tuned to r$^2$SCAN+rVV10 DFT calculations (Methods and SI Sec.~S1\,B\,1). The fine-tuned potential achieves mean absolute errors of 0.20~meV/atom in energy and 10.8~meV/\AA{} in force on a held-out test set of the three polymorphs at 800~K (SI Fig.~S2).
We simulated each polymorph in its experimental crystal structure at 1~bar with an isotropic barostat, which preserves the experimental cell shape to prevent framework interconversion while allowing Li ions to hop freely between sites.
The $\gamma$ polymorph is analyzed only up to 800~K because its \ce{PS4} tetrahedra begin to rearrange at 850~K (Methods and SI Fig.~S3e). 

Figure~\ref{fig:lps_stability}b shows the configurational entropy of Li, $S_\mathrm{config}^{(\mathrm{Li})}$, directly predicted by EntroVAMP models trained separately for each polymorph and temperature. Across the simulated temperature range, $S_\mathrm{config}^{(\mathrm{Li})}$ follows the order $\gamma \le \beta < \alpha$, remaining near zero in both $\gamma$ and $\beta$ below 350~K. It rises steeply between 350 and 425~K in $\alpha$, between 450 and 650~K in $\beta$, and above 725~K in $\gamma$, consistent with the onset of Li diffusion in each polymorph (SI Fig.~S3d). 
Although trained without crystallographic site labels, the learned Li states map onto distinct Wyckoff site families in each polymorph (SI Fig.~S13). In $\gamma$, the two learned states correspond to the $4b$ and $2a$ sites, whereas in $\beta$ the two low-temperature states correspond to the Li1 and Li2 ($8d$) sites and a third state present at 450~K maps onto the interstitial Li3 ($4c$) site.
In $\alpha$, three of the four states reside mainly on the $16h$, $8e$, and $4c$ sites (70--99\% of their time), and the fourth spends about two-thirds of its time near $16h$ sites and one-third near $8e$ sites.

Figure~\ref{fig:lps_stability}c plots the eigenvalues $\lambda_k$ of the Li transition matrix versus lag time at 450~K. The eigenvalue criterion selects these state counts ($K=2$ in $\gamma$, $3$ in $\beta$, and $4$ in $\alpha$) because training with $K+1$ states produces an unphysical negative eigenvalue $\lambda_{K+1}$ in each polymorph. 
The eigenvalues also give the timescale on which the Li atoms change state. In $\gamma$, $\lambda_2$ remains near unity across all plotted lags up to 1~ns, indicating that Li ions rarely change state within this window. In $\beta$ and $\alpha$, by contrast, $\lambda_2$ decays within about 1~ns and 40~ps, respectively, consistent with faster Li diffusion in $\alpha$ than in $\beta$.

Figure~\ref{fig:lps_stability}d compares the distributions of Li vibrational periods and duration-weighted Li state lifetimes from 300 to 800~K. At 300~K, the two barely overlap in any polymorph. As the temperature rises, the state lifetimes shorten and begin to overlap with the vibrational periods, so that vibrations and changes of state are no longer separated in time. Evaluating $S(T)$ directly as $S_\mathrm{vib}(T)+S_\mathrm{config}(T)$ at elevated temperatures may double-count diffusive Li motion that enters the vibrational spectrum, overestimating the entropies of $\beta$ and $\alpha$.
To avoid this high-temperature double counting, we evaluate the absolute reference entropy $S(T_0)=S_\mathrm{vib}(T_0)+S_\mathrm{config}(T_0)$ with EntroVAMP only at an anchor temperature $T_0$ low enough for vibrational and hopping timescales to remain separated. We then obtain the entropy at other temperatures by thermodynamic integration of the MD energy. At constant pressure the entropy changes with temperature as $T\,\mathrm{d}S=\mathrm{d}U+P\,\mathrm{d}V$, and at 1~bar the $P\,\mathrm{d}V$ term is negligible (SI Sec.~S4\,B\,3), so that
\begin{equation}
S(T)=S(T_0)+\int_{T_0}^{T}\frac{1}{T'}\frac{\mathrm{d}U}{\mathrm{d}T'}\,\mathrm{d}T',
\label{eq:ti}
\end{equation}
which requires no separation of vibration and configuration. 
The free energy then follows as $F(T)=U(T)-TS(T)$. Selecting $T_0$ involves a trade-off between timescale separation and configurational sampling, particularly in $\alpha$-\ce{Li3PS4}, where Li hops are already thermodynamically active at low temperature yet rare on nanosecond MD timescales. Lowering $T_0$ toward 300~K suppresses these rare hops and under-samples $S_\mathrm{config}$ within 5~ns trajectories (SI Fig.~S15), whereas raising $T_0$ to 375~K or above causes $S_\mathrm{vib}+S_\mathrm{config}$ to exceed the integral of Eq.~\eqref{eq:ti} as hopping and vibrational timescales overlap (SI Fig.~S16a). 
We therefore set $T_0=350$~K at which $S_\mathrm{vib}+S_\mathrm{config}$ agrees with Eq.~\eqref{eq:ti} across all three polymorphs, and integrate Eq.~\eqref{eq:ti} both above and below $T_0$ to evaluate $S(T)$ and $F(T)$ from 300 to 1000~K. 
Because single 5~ns trajectories do not exhaustively sample the rare Li hops, $S^{\alpha}_\mathrm{config}$ spans 0.10 to 0.43~J/mol$_\mathrm{atom}$/K across five independent MD runs with different velocity seeds. We therefore use the ensemble mean over these five runs with its sampling variance propagated. 

Figure~\ref{fig:lps_stability}e shows the resulting free energies of $\beta$ and $\alpha$ relative to $\gamma$, together with the measured transition ranges. The free-energy difference $F_\beta-F_\gamma$ crosses zero at $595\pm13$~K (uncertainties denote two standard errors throughout), close to the experimental $\gamma\rightarrow\beta$ transition range of 523--573~K. The inset of Fig.~\ref{fig:lps_stability}e plots $F_\alpha-F_\beta$ up to 1000~K. 
This difference crosses zero at 875~K, with the lower edge of its $2\sigma$ uncertainty band reaching zero at 680~K and spanning the experimental $\beta\rightarrow\alpha$ range of 723--773~K. 
Omitting $S_\mathrm{config}(T_0)$ shifts the $\beta\rightarrow\alpha$ crossing above 1000~K, whereas the $\gamma\rightarrow\beta$ crossing shifts by less than 1~K, as this transition is mostly driven by the vibrational entropy difference in the temperature range.
Between 700 and 1000~K, $F_\alpha$ and $F_\beta$ remain within 1~meV/atom of each other with a shallow relative slope; sub-meV/atom uncertainties from rare-hop MD sampling at $T_0$, numerical temperature integration, and DFT or MLIP errors can broaden the $\beta\rightarrow\alpha$ transition band.
Nevertheless, the framework still recovers the $\gamma\rightarrow\beta\rightarrow\alpha$ stability sequence close to experiments from \textit{ab initio} bottom-up modeling.

\begin{figure*}[t]
\centering
\includegraphics{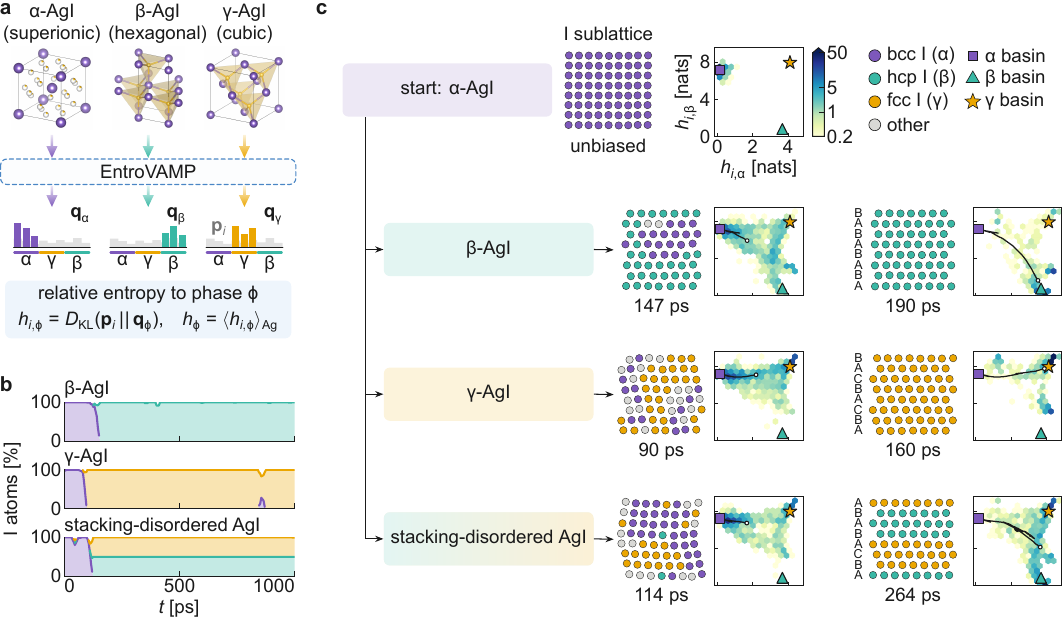}
\caption{\textbf{Enhanced sampling of phase transitions in AgI at 380~K.}
\textbf{a}, Unbiased 380~K trajectories of superionic $\alpha$-AgI, $\beta$-AgI, and $\gamma$-AgI are encoded by a single EntroVAMP model shared by the three phases. Averaging the Ag state probabilities over each trajectory gives the reference distribution $\mathbf{q}_\phi$ of phase $\phi$. During enhanced sampling, the trained model is applied to each new configuration and returns the state probabilities $\mathbf{p}_i$ of every Ag atom $i$. The relative entropy of $\mathbf{p}_i$ with respect to $\phi$ is $h_{i,\phi}=D_{\mathrm{KL}}(\mathbf{p}_i\,\|\,\mathbf{q}_\phi)$. Averaging over Ag atoms gives one collective variable (CV) per reference phase, $h_\phi=\langle h_{i,\phi}\rangle_{\mathrm{Ag}}$.
\textbf{b}, Percentages of iodine atoms classified as body-centered cubic (bcc, purple), hexagonal close-packed (hcp, teal), and face-centered cubic (fcc, gold) by adaptive common-neighbor analysis along the three enhanced-sampling runs of \textbf{c}, one row per run.
\textbf{c}, Well-tempered metadynamics biasing $h_\alpha$ drives the system out of the $\alpha$ basin.
The top row shows unbiased $\alpha$-AgI, and the three rows below show runs whose products are $\beta$-AgI, $\gamma$-AgI, and stacking-disordered AgI, each shown during and after the transition.
The snapshots display the iodine sublattice colored as in \textbf{b}, with letters indicating the stacking sequence of the eight close-packed layers.
$\alpha$-AgI is viewed down bcc $[001]$, and each biased run is viewed along the close-packed zone axis of its product phase.
Each hexbin plot shows the per-Ag distribution of $(h_{i,\alpha},h_{i,\beta})$ within $\pm2$~ps of the snapshot time, colored by the mean number of Ag atoms per bin and frame; the purple square, teal triangle, and gold star mark the reference $\alpha$, $\beta$, and $\gamma$ basins, and the black curve traces the mean $(h_\alpha, h_\beta)$ trajectory up to the snapshot time (open circle).}
\label{fig:agi_es}
\end{figure*}

\subsection{Entropy-driven enhanced sampling}\label{subsec:es}
Beyond estimating the configurational entropy, we next show that the information entropy from EntroVAMP can serve as a collective variable (CV) for enhanced sampling.
Many processes of interest, such as nucleation and solid--solid phase transitions, require crossing free-energy barriers on timescales inaccessible to unbiased MD simulations.
Enhanced sampling accelerates such transitions by adding a bias along a few CVs that lowers the free-energy barriers~\cite{laioEscapingFreeenergyMinima2002, barducciWTMetaD2008}.
The efficiency of enhanced sampling depends on a CV that distinguishes the metastable states and follows the slow motion along the transition pathway, and entropy has served as a CV for phase transitions~\cite{piaggiEntropyEnthalpy2017, piaggiOrientationalEntropy2018,  amodeoNanoparticleFreezing2020, mendelsSearchingEntropicallyStabilized2017,  liInformationEntropyGeneralPurpose2026a}.
Prior entropy CVs are either structural entropies computed from pair correlations~\cite{piaggiEntropyEnthalpy2017, piaggiOrientationalEntropy2018, mendelsSearchingEntropicallyStabilized2017, amodeoNanoparticleFreezing2020} or information entropies that compare each atomic environment with a reference set in a predefined descriptor space~\cite{liInformationEntropyGeneralPurpose2026a}.
We instead compute the information entropy from each atom's soft probabilities over the coarse-grained states that EntroVAMP learns for each phase.
The EntroVAMP network, the softmax and the entropy are all differentiable functions, so the entropy CV is also differentiable.

The CV must be defined consistently across phases and distinguish the source phase from its products, so we train one joint EntroVAMP model on unbiased trajectories from all phases involved.
The joint model has one block $\mathcal{B}(\phi)$ of states per phase $\phi$, with as many states as the eigenvalue criterion selected when EntroVAMP was trained on that phase alone (e.g., for AgI, five for $\alpha$, five for $\gamma$ and four for $\beta$ give $K=14$ Ag states).
For every atom $i$, the model returns a probability vector $\mathbf{p}_i=(p_i(1),\dots,p_i(K))$ that sums to one, so that $P_i(\phi)=\sum_{s\in\mathcal{B}(\phi)}p_i(s)$ is the probability that the model assigns the environment of atom $i$ to phase $\phi$.
A cross-entropy term, $-\ln P_i(\phi)$ averaged over the trajectory of phase $\phi$, ties each block to its phase, while the VAMP-2 score, evaluated on each phase's trajectory separately, determines how the states of a block divide that phase's environments (see Methods, Eq.~\eqref{eq:es_loss}).
An atom in a bulk phase then carries its probability on the block of that phase, whereas an atom in a transforming region carries probability on more than one block.
The simplest CV is the mean $h$ over atoms of the per-atom information entropy $h_i=-\sum_s p_i(s)\ln p_i(s)$, which is near zero when every atom sits firmly in one state and larger when the probabilities spread over several states. 
The CV $h$ therefore distinguishes disordered from ordered phases.

We apply EntroVAMP to drive the solid--solid phase transitions of AgI at 380~K, starting from the superionic $\alpha$ phase.
Averaging the Ag state probabilities over the unbiased trajectory of each phase gives the reference distribution $\mathbf{q}_\phi$ of phase $\phi\in\{\alpha,\beta,\gamma\}$ (Fig.~\ref{fig:agi_es}a).
The ordered phases can have similar entropies, as $\beta$- and $\gamma$-AgI do, so $h$ alone does not separate them.
We therefore measure how far each Ag atom's local environment is from phase $\phi$ by the relative entropy $h_{i,\phi}=D_{\mathrm{KL}}(\mathbf{p}_i\|\mathbf{q}_\phi)$, the Kullback--Leibler divergence of its probabilities from $\mathbf q_\phi$~\cite{gobboRelativeEntropy2018}.
The three phases of AgI are well separated in the plane of $h_{i,\alpha}$ and $h_{i,\beta}$ (Fig.~\ref{fig:agi_es}c).
The CV is the mean $h_\alpha$ over Ag atoms, which drives the system out of $\alpha$ without selecting the product.
The biased runs turn the bcc iodine sublattice of $\alpha$-AgI into the close-packed sublattices of $\beta$-AgI, $\gamma$-AgI and stacking-disordered AgI (Fig.~\ref{fig:agi_es}b,c).
Across 20 independent runs biasing $h_\alpha$, the first conversion yields stacking-disordered AgI 18 times and $\beta$- and $\gamma$-AgI once each.
The per-Ag relative entropies are first centered on the $\alpha$ basin, spread out during the transition, and, for $\beta$- and $\gamma$-AgI, finally gather around the product basin (Fig.~\ref{fig:agi_es}c).
Below 420~K, $\beta$-AgI is the stable phase and $\gamma$-AgI is metastable, but cooling $\alpha$-AgI in experiments yields stacking-disordered AgI~\cite{smithStackingDisorderDesign2019}.
Although EntroVAMP is trained only on $\alpha$-, $\beta$- and $\gamma$-AgI, the biased runs also sample stacking-disordered AgI, with different stacking sequences (Fig.~\ref{fig:agi_es}c and Extended Data Fig.~\ref{edfig:agi_sd_gallery_gnn}).

In the enhanced sampling above, two networks read the same structure. 
The GET encoder of EntroVAMP feeds the VAMPNet head that provides the CV, while MACE-MH-1 evaluates the energy and forces for MD simulations.
Both compress each local atomic environment into a latent representation, so we asked whether the frozen latent of the potential can replace the trained GNN encoder.

We trained only a VAMPNet head on the frozen MACE-MH-1 features.
Biasing $h_\alpha$ built from this head again drives $\alpha$-AgI out of its basin and yields $\beta$-, $\gamma$-, and stacking-disordered AgI (Extended Data Fig.~\ref{edfig:agi_es_latent} and SI Table~S7).
The pretrained MACE-MH-1 latent therefore already distinguishes the Ag environments of the three phases.
This agreement between two encoders trained on different signals is the behavior expected if atomistic representations converge toward a shared description of local structure \cite{huh_position_2024, li_platonic_2026}, a point we return to in the Discussion.
In practice, the CV can then be read from the same network evaluation that gives the energy and forces.

\section{Discussion}

EntroVAMP evaluates the configurational entropy from a single MD trajectory, without a reference lattice or a hand-picked structural observable.
The configurational entropy is computed as the Shannon entropy of the distribution over coarse-grained states (Eq.~\eqref{eq:S_split}).
Although the Shannon entropy can be evaluated for any partition of atomic environments into coarse-grained states \cite{jaynes_information_1957}, it equals the thermodynamic configurational entropy only when the system equilibrates vibrationally within each state and leaves it by slow hops.
EntroVAMP learns this separation from dynamics rather than an arbitrary static site partition.
An equivariant GNN encodes local atomic environments \cite{batznerNequIP2022,  sheriffChemicalMotif2024}, and the VAMP objective selects from the resulting latent representations the coarse-grained states whose transitions are slow compared with vibrations.
The configurational and vibrational entropies are then separated by timescale, which is the condition for the configurational entropy to be well defined \cite{ceder_derivation_1993}.

EntroVAMP provides a probabilistic description of each phase, and of the intermediate structures between phases, as a distribution over the same coarse-grained states.
A phase transition can then be viewed as a flow of probability from the distribution of the source phase to that of the product phase, with progress measured by the relative entropy of the per-atom state probabilities with respect to a reference phase.
The learned states supply collective variables for enhanced sampling with three properties: (1) they are built from soft state assignments and are differentiable with respect to the atomic coordinates for direct biasing; (2) they are defined consistently across phases because a single model encodes all of them; and (3) they presuppose no order parameter of the product phase.
For instance, biasing only the relative entropy with respect to the source phase drives $\alpha$-AgI into $\beta$-AgI, $\gamma$-AgI and stacking-disordered AgI (Fig.~\ref{fig:agi_es}c), the last being a product absent from the training set but observed experimentally upon cooling \cite{smithStackingDisorderDesign2019}.
Prior work on entropy-based CVs computes the entropy from hand-chosen structural quantities, such as pair correlations \cite{piaggiEntropyEnthalpy2017, piaggiOrientationalEntropy2018} or static descriptor distributions \cite{liInformationEntropyGeneralPurpose2026a}.
The information-entropy CVs of EntroVAMP are instead built on states learned from the slow dynamics within each phase.
Another approach learns the CV itself from the dynamics of the transition \cite{bonatiDeepTICA2021}, which requires trajectories in which the transition has already been sampled.
In contrast, EntroVAMP needs only unbiased trajectories of the individual phases.
A natural next step is to converge free-energy surfaces along these information-entropy CVs to obtain transition barriers and nucleation rates.

\begin{figure}[t]
\centering
\includegraphics[width=0.75\columnwidth]{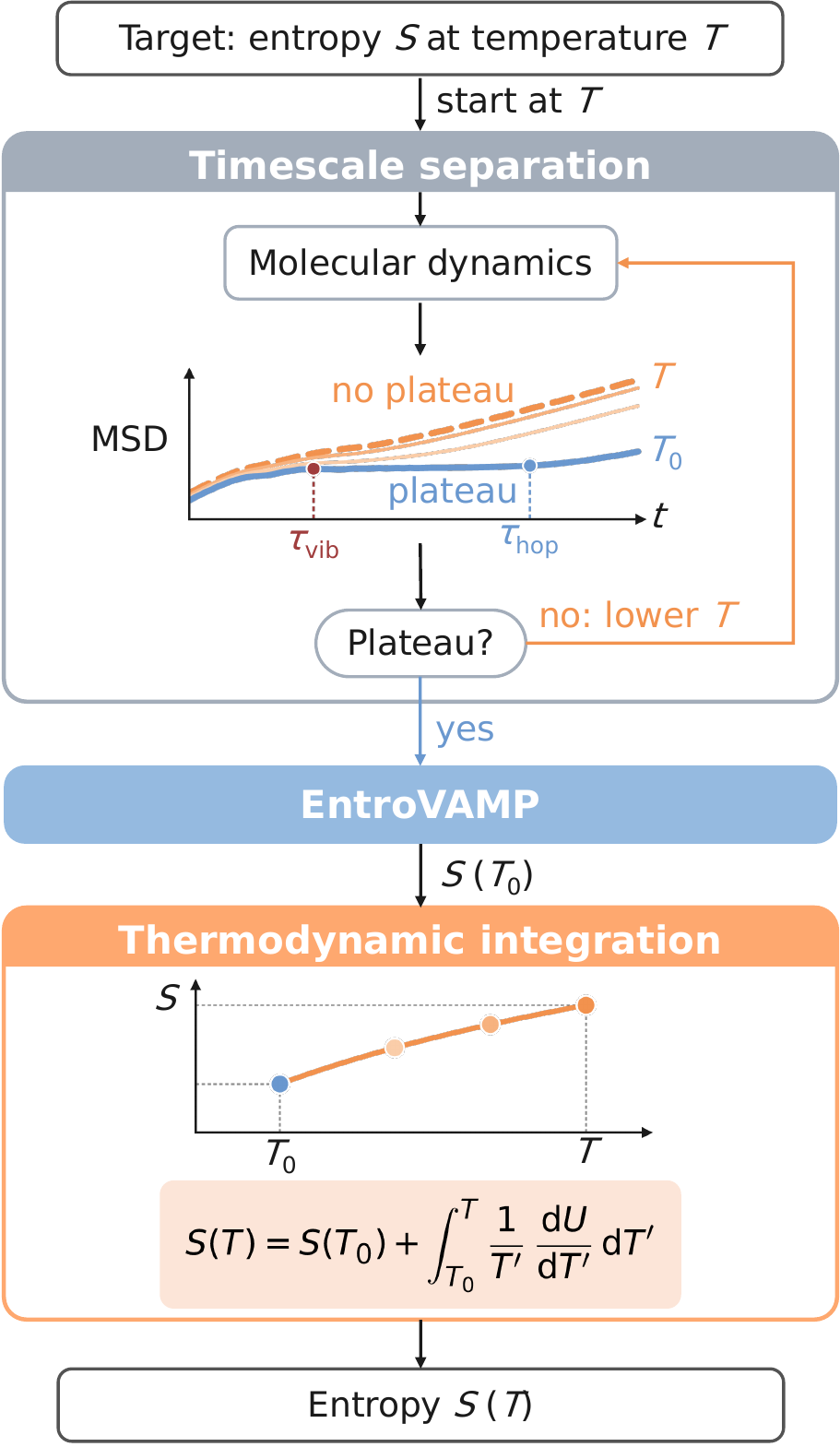}
\caption{\textbf{Workflow for computing entropy at a target temperature $T$.} The MD temperature is lowered from $T$ until the mean squared displacement (MSD) exhibits a plateau between $\tau_{\mathrm{vib}}$ and $\tau_{\mathrm{hop}}$, indicating timescale separation at an anchor temperature $T_0$. EntroVAMP yields the reference entropy $S(T_0)$, and thermodynamic integration of the internal energy $U$ from $T_0$ to $T$ gives $S(T)$.}
\label{fig:workflow}
\end{figure}

We note two caveats in applying EntroVAMP for entropy estimation.
First, the method relies on separating vibrational and hopping timescales, and this separation narrows as temperature rises (Fig.~\ref{fig:workflow}).
When atoms hop within a few vibrational periods at high temperatures, the spectral gap that determines $K$ closes, part of the diffusive hopping motion enters the vibrational spectrum, and $S_{\mathrm{vib}}$ and $S_{\mathrm{config}}$ can no longer be cleanly separated, which can lead to double counting.
The remedy is to compute the absolute entropy at a lower temperature $T_0$, where the timescales are well separated, directly with EntroVAMP and to use it as the reference entropy (Fig.~\ref{fig:workflow}).
In contrast, conventional thermodynamic integration obtains the reference by integrating along a path from a model system of known free energy (e.g., Einstein or Debye crystals \cite{freitas_nonequilibrium_2016, cheng_computing_2018}).
The two approaches differ only in this reference and share the subsequent thermodynamic integration over temperature from $T_0$ to $T$.
Second, a single unbiased MD trajectory samples the hops of mobile species but not the slow substitutional exchanges of a chemically disordered sublattice \cite{peng_atomistic_2026}.
Hybrid Monte Carlo/MD simulations~\cite{sadigh_hybrid_2012} can sample those exchanges, though they may require substantially longer trajectories to reach ergodicity.
This suggests a natural division of labor: Monte Carlo handles the combinatorics of substitutional site occupation, whereas EntroVAMP captures off-lattice disorder (e.g., vacancies and Frenkel defects that displace surrounding atoms) directly from the MD trajectory whenever those defects are sampled.

The dynamical information that defines configurational states is encoded directly in the atomistic latent space. This perspective connects to recent efforts to characterize atomistic latent and descriptor spaces \cite{schwalbe-koda_model-free_2025} and the Platonic representation hypothesis, which posits that models trained with different architectures and objectives converge toward a shared representation of physical reality \cite{huh_position_2024}.
In atomistic modeling, latent-space comparisons show that foundation potentials align onto a common geometry that preserves chemical periodicity and structural invariants \cite{li_platonic_2026}, with high-performing models converging on in-distribution structures \cite{edamadaka_universally_2025}.
EntroVAMP organizes the latent space using a complementary signal (i.e., the time-lagged dynamics on the PES), yet converges on coarse-grained states that reflect the underlying structural motifs.
If this dynamical partition is governed by the physical timescales of the material rather than by the model architecture, the resulting $S_{\mathrm{config}}$ should not depend on the encoder, a prediction that can be tested by training VAMPNet heads on the frozen latents of different potentials.
By reproducing the calorimetric transition entropy of AgI and the observed polymorph sequence of \ce{Li3PS4} (Fig.~\ref{fig:lps_stability}) without fitting to empirical entropy data, EntroVAMP complements model-to-model representational alignment \cite{li_platonic_2026, edamadaka_universally_2025} with a direct thermodynamic test of whether an atomistic latent space resolves physically relevant states.

Looking ahead, as MLIPs increasingly provide accurate energies and forces and make accurate MD routine across chemistries, we envision EntroVAMP as their counterpart for accurate entropies and free energies from atomistic simulations.
A series of MD trajectories for each phase yields the internal energy at every temperature and the vibrational and configurational entropy at the reference temperature, all at one level of theory and without a reference lattice or a hand-chosen descriptor.
EntroVAMP thus supplies the configurational term that $0$~K convex hulls omit for disordered, superionic, and metastable phases, and could make finite-temperature predictions of phase stability and synthesizability as routine as $0$~K screening is today.
Furthermore, the entropy-based CV can be read from a small VAMPNet head using the same latent representation that drives the MD, making enhanced sampling as simple to set up as MD itself.
Being both effective and simple, enhanced sampling with EntroVAMP CVs could aid the exploration of transition pathways, the identification of metastable phases, and the computation of free-energy barriers.

\section*{Methods}

\noindent\textbf{MD simulations.}
For AgI, we performed MD simulations in LAMMPS~\cite{Thompson2022LAMMPS} using the MatPES r$^2$SCAN (\texttt{matpes\_r2scan}) head of the MACE-MH-1 foundation potential~\cite{Batatia2022MACE, Batatia2025MACEMH1, kaplan_foundational_2025} with a time step of 1~fs. Temperature was controlled using a Nos\'e--Hoover chain thermostat~\cite{Martyna1992NHC}, and pressure during $NPT$ equilibration was controlled using a Nos\'e--Hoover barostat with the Martyna--Tobias--Klein correction~\cite{Martyna1994MTK}. The simulation supercells for the $\alpha$ (superionic), $\beta$, and $\gamma$ phases of AgI contained 512, 432, and 512 atoms, respectively. All three phases were equilibrated for 200~ps in the $NPT$ ensemble at 400~K and 1~bar. For each phase, we selected the snapshot whose density was closest to the $NPT$ mean to initialize a 2~ns $NVT$ production run at 400~K, from which we evaluated the configurational and vibrational entropies. 
For \ce{Li3PS4}, we performed MD simulations in LAMMPS~\cite{Thompson2022LAMMPS} in double precision with a 1~fs time step, using the MACE-MH-1 potential fine-tuned to r$^2$SCAN+rVV10 (SI Sec.~S1\,B\,1). Each polymorph was simulated in a 512-atom supercell constructed from its experimental crystal structure: $\gamma$ ($Pmn2_1$)~\cite{homma_crystal_2011}, $\beta$ ($Pnma$), and $\alpha$ ($Cmcm$)~\cite{kaup_impact_2020}. Temperature and pressure (1~bar) were controlled using a Nos\'e--Hoover thermostat and barostat with damping times of 0.1 and 1.0~ps, respectively. The barostat was isotropic, allowing volume fluctuations while preserving the cell shape. After energy minimization and 200~ps of $NPT$ equilibration, we collected a 5~ns $NPT$ production trajectory at each temperature, saving atomic positions and velocities every 6~fs. Simulations were conducted from 300 to 400~K in steps of 12.5~K, from 400 to 800~K in steps of 25~K and from 850 to 1000~K in steps of 50~K. The $\gamma$ polymorph was analyzed up to 800~K because its P--P distances begin to change at 850~K as the \ce{PS4} framework rearranges (SI Fig.~S3e). Further details are provided in SI Sec.~S1\,B\,2 and SI Table~S2.

\noindent\textbf{Vibrational entropy.}
We evaluated the vibrational entropy $S_{\mathrm{vib}}$ from the vibrational density of states (VDOS) of the MD trajectories \cite{lin_two-phase_2003}. For a system of $N$ atoms, the mass-weighted velocity autocorrelation function (VACF) is defined as
\begin{equation}
  C(t) = \sum_{i=1}^{N}
         m_i\, \langle \mathbf{v}_i(t_0) \cdot \mathbf{v}_i(t_0 + t) \rangle ,
\end{equation}
where $\mathbf{v}_i(t)$ is the velocity of atom $i$ (with the center-of-mass velocity of the cell removed), $m_i$ is its mass, and $\langle \cdots \rangle$ denotes the ensemble average over time origins $t_0$. Mass weighting ensures that each vibrational mode carries the same thermal energy $k_BT$ in a multicomponent cell. The VDOS $g(\omega)$ is obtained from the Fourier transform of $C(t)$,
\begin{equation}
  g(\omega) = \int_{-\infty}^{\infty} C(t)\, e^{-i\omega t}\, \mathrm{d}t ,
\end{equation}
and is normalized such that $\int_{0}^{\infty} g(\omega)\, \mathrm{d}\omega = 3N$, corresponding to the total number of vibrational modes.
Each vibrational mode at angular frequency $\omega$ is treated as a classical harmonic oscillator, consistent with the classical MD simulations used to evaluate $U$, giving the vibrational entropy
\begin{equation}
S_{\mathrm{vib}} = k_B \int_{0}^{\infty} g(\omega) \left[ 1 - \ln\!\left(\frac{\hbar\omega}{k_BT}\right) \right] \mathrm{d}\omega ,
\end{equation}
\noindent where $\hbar$ is the reduced Planck constant, $k_B$ is the Boltzmann constant, and $T$ is the temperature.

\noindent\textbf{EntroVAMP implementation.}
Consider an MD trajectory of $M$ frames $\{\mathbf{x}_1,\dots,\mathbf{x}_M\}$, where $\mathbf{x}_t$ denotes the coordinates of all $N$ atoms in frame $t$. For a lag time $\tau$, the trajectory provides $M-\tau$ time-lagged pairs $\{(\mathbf{x}_t,\mathbf{x}_{t+\tau})\}_{t=1}^{M-\tau}$. The supercell contains species $e \in \{1,\dots,E\}$ with $N_e$ atoms of species $e$, for a total of $N=\sum_{e=1}^{E} N_e$ atoms. We omit the time index $t$ from per-atom quantities ($\mathbf{p}_i$, $s_i$, and $\mathbf{n}_i$) and write it explicitly as a subscript (e.g., $\mathbf{p}_{i,t}$) only where needed.

The encoder stacks $L$ message-passing layers of the BAMBOO graph equivariant transformer (GET)~\cite{bamboo_2025}, constructing the neighbor graph under the minimum-image convention for periodic supercells. The network takes only the atomic numbers $Z_i$ and interatomic vectors $\mathbf{r}_{ij}$ within a cutoff radius $r_c$ as inputs, without hand-crafted descriptors. Each layer $l \in \{0,\dots,L-1\}$ maintains an $\mathrm{SE}(3)$-invariant scalar feature $\mathbf{h}_i^{l}$ and an $\mathrm{SE}(3)$-equivariant vector feature $\vec{\mathbf{v}}_i^{l}$, updating both over the local neighborhood $\mathcal{N}(i)=\{\,j: d_{ij}=\lVert\mathbf{r}_{ij}\rVert<r_c\,\}$:
\begin{equation}
  \big(\mathbf{h}_i^{l+1}, \vec{\mathbf{v}}_i^{l+1}\big) = \mathrm{GET}^{l}\Big(\mathbf{h}_i^{l}, \vec{\mathbf{v}}_i^{l}, \big\{\mathbf{h}_j^{l}, \vec{\mathbf{v}}_j^{l}, \mathbf{r}_{ij}\big\}_{j\in\mathcal{N}(i)}\Big),
  \label{eq:gnn}
\end{equation}
initialized from $\mathbf{h}_i^{0}=\mathrm{embed}(Z_i)$ and $\vec{\mathbf{v}}_i^{0}=\mathbf{0}$. Because the vector channel propagates bond directions, the encoder resolves angular geometry in addition to pairwise distances. The final layer outputs the invariant embedding $\mathbf{h}_i=\mathbf{h}_i^{L}$ of the local environment of atom $i$. A single shared encoder feeds species-specific heads $\mathrm{VAMPNet}^{(e)}$, each consisting of a linear layer followed by a $K_e$-way softmax, where $K_e$ is the number of coarse-grained states for species $e$. For each frame $\mathbf{x}_t$, the head maps $\mathbf{h}_i$ of atom $i$ (of species $e$) to a soft probability vector over its $K_e$ states $s \in \{1,\dots,K_e\}$:
\begin{equation}
  \mathbf{p}_{i} = \mathrm{VAMPNet}^{(e)}(\mathbf{h}_{i}) \in \mathbb{R}^{K_e}, \qquad i = 1,\dots,N_e .
  \label{eq:soft_assign}
\end{equation}

We train these state assignments to resolve slow configurational dynamics using a per-species VAMP-2 objective. For each species $e$, the $N_e$ chemically identical atoms share one head and are treated as trajectories of a common single-site Markov chain, with expectations averaged over all atoms and time-lagged frame pairs:
\begin{equation}
  \langle\,\cdot\,\rangle_e \equiv \frac{1}{(M-\tau)N_e}\sum_{t=1}^{M-\tau}\sum_{i=1}^{N_e}(\,\cdot\,) .
\end{equation}
With mean vectors $\boldsymbol{\pi}_0^{(e)} = \langle\mathbf{p}_{i,t}\rangle_e$ and $\boldsymbol{\pi}_1^{(e)} = \langle\mathbf{p}_{i,t+\tau}\rangle_e$, the mean-subtracted instantaneous and time-lagged covariance matrices are
\begin{align}
  \bar{\mathbf{C}}^{(e)}_{00} &= \langle\mathbf{p}_{i,t}\mathbf{p}_{i,t}^{\top}\rangle_e - \boldsymbol{\pi}_0^{(e)}\boldsymbol{\pi}_0^{(e)\top}, \nonumber\\
  \bar{\mathbf{C}}^{(e)}_{11} &= \langle\mathbf{p}_{i,t+\tau}\mathbf{p}_{i,t+\tau}^{\top}\rangle_e - \boldsymbol{\pi}_1^{(e)}\boldsymbol{\pi}_1^{(e)\top}, \label{eq:covariances}\\
  \bar{\mathbf{C}}^{(e)}_{01} &= \langle\mathbf{p}_{i,t}\mathbf{p}_{i,t+\tau}^{\top}\rangle_e - \boldsymbol{\pi}_0^{(e)}\boldsymbol{\pi}_1^{(e)\top}. \nonumber
\end{align}
The per-species VAMP-2 loss is defined as the negative of the VAMP-2 score (the squared Frobenius norm of the mean-free Koopman matrix plus one for the trivial stationary mode):
\begin{equation}
  \mathcal{L}_{\mathrm{vamp2}}^{(e)} = -\big\lVert (\bar{\mathbf{C}}^{(e)}_{00})^{-1/2}\,\bar{\mathbf{C}}^{(e)}_{01}\,(\bar{\mathbf{C}}^{(e)}_{11})^{-1/2} \big\rVert_F^2 - 1 .
  \label{eq:vamp2}
\end{equation}
Here, $(\bar{\mathbf{C}}^{(e)}_{00})^{-1/2}\,\bar{\mathbf{C}}^{(e)}_{01}\,(\bar{\mathbf{C}}^{(e)}_{11})^{-1/2}$ is the mean-free Koopman matrix of the coarse-grained dynamics, whose squared Frobenius norm equals the sum of squared singular values $\sum_k (\sigma^{(e)}_k)^2$, where each $\sigma^{(e)}_k \in [0,1]$ is the time-lagged correlation of a collective mode of the state assignment. Because each soft assignment sums to unity, the stationary mode is identically conserved and contributes a trivial singular value $\sigma_0 = 1$; mean subtraction removes this mode from the Frobenius norm, and the constant term in Eq.~\eqref{eq:vamp2} restores its contribution. A singular value near unity corresponds to a mode that persists over the lag time $\tau$. Minimizing $\mathcal{L}_{\mathrm{vamp2}}^{(e)}$ therefore drives the network to resolve the slowest configurational transitions while averaging over faster intra-basin vibrations. In general, the shared encoder and all species heads can be trained jointly by minimizing the species-weighted loss
\begin{equation}
  \mathcal{L} = \sum_{e} w_e\,\mathcal{L}_{\mathrm{vamp2}}^{(e)}, \qquad w_e = \frac{N_e}{N} . 
  \label{eq:loss}
\end{equation}
To scan the number of states $K_e$ independently for each species rather than over combinatorial tuples $\{K_e\}$, in this work we trained a separate model for each active species $e$, assigning $K_e$ states to species $e$ and a single state ($K_{e'}=1$) to all other species.

The lag time $\tau$ must be shorter than $\tau_{\mathrm{hop}}$ but cannot be arbitrarily small: as $\tau\to0$, the time-lagged correlations of all assignments approach unity and the VAMP-2 score no longer separates slow configurational transitions from vibrations. We used $\tau=300$~fs for AgI and $180$~fs for \ce{Li3PS4}. 
We determine $K_e$ from the spectrum of the transition matrix $\mathbf{T}^{(e)}(\tau)=(\mathbf{C}^{(e)}_{00})^{-1}\mathbf{C}^{(e)}_{01}$, constructed from the uncentered correlation matrices $\mathbf{C}^{(e)}_{00}=\langle\mathbf{p}_{i,t}\mathbf{p}_{i,t}^{\top}\rangle_e$ and $\mathbf{C}^{(e)}_{01}=\langle\mathbf{p}_{i,t}\mathbf{p}_{i,t+\tau}^{\top}\rangle_e$ with negative entries set to zero and rows renormalized. Scanning candidate state counts, we select the largest $K_e$ for which a model trained with $K_e+1$ states yields a $(K_e+1)$-th eigenvalue $\lambda_{K_e+1}$ that is either negative at some lag time or separated from $\lambda_{K_e}$ by a large spectral gap (a negative eigenvalue corresponds to an unphysical process, because it cannot be assigned a relaxation timescale). For AgI and \ce{Li3PS4}, incrementing the state count to $K_e+1$ produces a negative $\lambda_{K_e+1}$ or a large spectral gap, as in the three-well potential (Fig.~\ref{fig:toy_agi}b,c), whereas for alanine dipeptide an eight-state model exhibits a clear spectral gap after the fourth eigenvalue (SI Fig.~S4). A species whose atoms remain in a single local environment has $K_e=1$ and zero configurational entropy.

\noindent\textbf{Configurational entropy estimation.}
During inference, the local environment of each atom $i$ of species $e$ in a given frame is assigned to its most probable state, $s_i = \arg\max_s p_i(s)$. Let $\hat{p}_i(s)$ denote the empirical state probability of atom $i$ accumulated over the $M$ analyzed frames, forming the vector $\hat{\mathbf{p}}_i$, and let $H(\mathbf{p}) = -\sum_s p(s)\ln p(s)$ denote the Shannon entropy of a discrete distribution $\mathbf{p}$. 
For each species $e$, every frame carries a population vector $(N_e(1),\dots, N_e(K_e))$ counting how many of its $N_e$ atoms occupy each state. For a liquid or a superionic sublattice at elevated temperature, atoms exchange freely within the observation window, making all $N_e!/\prod_{s} N_e(s)!$ arrangements accessible. The atoms are then statistically indistinguishable, and their empirical marginals agree ($\hat{\mathbf{p}}_i \approx \hat{\mathbf{p}}_j$ for any pair of atoms $i,j$) up to finite-sampling noise. As $N_e \to \infty$, the configurational entropy of such an ergodic species converges to the mean per-atom entropy $S_\mathrm{atom}^{(e)}$ (Eq.~\eqref{eq:satom} evaluated over the $N_e$ atoms of species $e$).

For a non-ergodic species~\cite{palmerBrokenErgodicity1982, mauroContinuouslyBrokenErgodicity2007}, such as Li in \ce{Li3PS4} below the onset of Li diffusion in each polymorph, atoms occasionally hop but do not exchange freely within the observation window. Only a small fraction of the $N_e!/\prod_s N_e(s)!$ arrangements are realized. The states of neighboring atoms remain correlated, and $S_\mathrm{atom}^{(e)}$ overestimates the configurational entropy. 
We therefore approximate the configurational entropy with the conditional entropy $S_\mathrm{loc}^{(e)}$ (Eq.~\eqref{eq:sloc_main} evaluated over the $N_e$ atoms of species $e$), which conditions on the state counts of the first-neighbor shell. 
This first-shell conditioning plays a role analogous to the pair and cluster corrections in mutual-information~\cite{killianExtractionConfigurationalEntropy2007} and cluster-variation~\cite{sanchezGeneralizedClusterDescription1984, fontaineClusterApproachOrderDisorder1994} expansions, but conditions on the entire first shell at once and requires no fixed reference lattice. In each frame, the count vector $\mathbf{n}_{i} = (n_i(1),\dots,n_i(K_e))$ tallies the states of the $m$ nearest same-species neighbors of atom $i$, where $n_i(s)$ is the number of those neighbors in state $s$ ($\sum_s n_i(s) = m$). The per-atom conditional entropy is evaluated as
\begin{equation}
  H\big(s_i \mid \mathbf{n}_i\big) = -\frac{1}{M}\sum_{t=1}^{M}\sum_{s} \hat{p}_i(s\mid\mathbf{n}_{i,t})\ln \hat{p}_i(s\mid\mathbf{n}_{i,t}),
  \label{eq:cond_entropy}
\end{equation}
where $\hat{p}_i(s\mid\mathbf{n}_i)$ is the fraction of frames with count vector $\mathbf{n}_i$ in which atom $i$ occupies state $s$. Averaging over all frames weights each count vector by its empirical frequency, yielding the conditional Shannon entropy of $s_i$ given $\mathbf{n}_i$. 
Because conditioning cannot increase entropy, $S_\mathrm{loc}^{(e)} \le S_\mathrm{atom}^{(e)}$; when $m=0$, $\mathbf{n}_i$ is empty, $H(s_i\mid\mathbf{n}_i)=H(s_i)$, and $S_\mathrm{loc}^{(e)}=S_\mathrm{atom}^{(e)}$. 
For a non-ergodic species, we set $m$ to cover the crystallographic first-shell neighbors because each atom retains its neighbors over the trajectory; conditioning on this shell removes the mutual information shared with them. For an ergodic species, where neighbors exchange too rapidly to sustain lasting correlations, $S_\mathrm{config}^{(e)} = S_\mathrm{atom}^{(e)}$ ($m=0$).
We select between $S_\mathrm{atom}^{(e)}$ and $S_\mathrm{loc}^{(e)}$ by testing whether all atoms of species $e$ share the same empirical state distribution $\hat{\mathbf{p}}_i$ (with $S_\mathrm{loc}^{(\mathrm{Li})}$ retained across all temperatures in \ce{Li3PS4} for continuity; SI Sec.~S2\,A and SI Table~S5). The total configurational entropy is the composition-weighted sum over all species:
\begin{equation}
  S_\mathrm{config} = \sum_e w_e\, S_\mathrm{config}^{(e)} .
  \label{eq:sconfig}
\end{equation}

\noindent\textbf{Enhanced sampling.}
The soft state probabilities produced by EntroVAMP can be used to construct CVs that drive phase transitions in enhanced-sampling simulations. Because a CV must be defined consistently across the phases and distinguish the source phase from its products, we train a single joint EntroVAMP model on unbiased trajectories of all phases involved. The total number of states in the joint network is $K_e=\sum_\phi K_\phi$, where $K_\phi$ is the number of states selected by the eigenvalue criterion for the single-phase model of phase $\phi$ in the configurational entropy analysis. The $K_e$ states are partitioned into disjoint blocks $\mathcal{B}(\phi)$ of size $K_\phi$, one per phase $\phi$. For AgI, across the superionic $\alpha$ phase (bcc iodine sublattice), the $\gamma$ phase (fcc iodine sublattice), and the $\beta$ phase (hcp iodine sublattice), the $K_\mathrm{Ag}=14$ Ag states comprise five $\alpha$ states ($\mathcal{B}(\alpha)=\{1,\dots,5\}$), five $\gamma$ states ($\mathcal{B}(\gamma)=\{6,\dots,10\}$), and four $\beta$ states ($\mathcal{B}(\beta)=\{11,\dots,14\}$). The total probability mass within a block, $P_{i}(\phi) = \sum_{s\in\mathcal{B}(\phi)}p_{i}(s)$, serves as a smooth indicator of the phase to which the local environment of atom $i$ belongs. We enforce this phase-to-block correspondence by adding a cross-entropy (CE) term on the block probability mass to the training loss, with the VAMP-2 loss evaluated separately on the trajectory of each phase:
\begin{align}
  \mathcal{L} &= \mathcal{L}_\mathrm{vamp2} + \lambda_\mathrm{CE}\,\mathcal{L}_\mathrm{CE}, \nonumber\\
  \mathcal{L}_\mathrm{CE} &= -\frac{1}{MN_e}\sum_{t=1}^{M}\sum_{i=1}^{N_e} \ln P_{i,t}(\phi) ,
  \label{eq:es_loss}
\end{align}
where $\phi$ is the phase label of the trajectory from which frame $t$ is drawn, $\lambda_\mathrm{CE}=1$, and $N_e$ is the number of atoms of the species $e$ whose states are partitioned into phase blocks. Because $\mathcal{L}_\mathrm{CE}$ is the negative log-likelihood of classifying each atom's local environment into its parent phase, the CE loss separates phases across blocks, whereas the VAMP-2 loss resolves slow, coarse-grained states within each phase.

Because biased simulations require a CV differentiable with respect to atomic coordinates, CVs are constructed from the soft probabilities $\mathbf{p}_i$ in the current frame rather than from discrete $\arg\max$ state assignments, and biased using well-tempered metadynamics~\cite{barducciWTMetaD2008}. The simplest CV is the mean per-atom information entropy $h=N_e^{-1}\sum_{i=1}^{N_e} h_i$, with $h_i=H(\mathbf{p}_i)$, which separates a disordered phase from an ordered one. 
For polymorphic transitions involving multiple crystalline phases (e.g., AgI), competing ordered polymorphs ($\beta$- and $\gamma$-AgI) need not differ in information entropy, so $h$ alone may not distinguish them. Following the strategy of comparing each local environment against a reference ensemble~\cite{liInformationEntropyGeneralPurpose2026a}, we instead define a phase-anchored CV using the relative entropy (Kullback--Leibler divergence) between $\mathbf{p}_i$ and the reference distribution of phase $\phi$:
\begin{align}
  h_\phi &= \frac{1}{N_e}\sum_{i=1}^{N_e} h_{i,\phi}, \nonumber\\
  h_{i,\phi} &= -\sum_{s} p_i(s)\ln\frac{q_\phi^{\epsilon}(s)}{p_i(s)}
           = D_\mathrm{KL}\big(\mathbf{p}_i \,\big\|\, \mathbf{q}_\phi^{\epsilon}\big) ,
  \label{eq:cv_kl}
\end{align}
where $\mathbf{q}_\phi^{\epsilon}$ is the mean state distribution of reference phase $\phi$ over its unbiased trajectory, floored at a small constant $\epsilon$ on states outside $\mathcal{B}(\phi)$ and renormalized so that $D_\mathrm{KL}$ remains finite. The per-atom divergence $h_{i,\phi}$ is small when $\mathbf{p}_i$ resembles $\mathbf{q}_\phi^{\epsilon}$ and grows steeply as probability weight shifts onto states outside $\mathcal{B}(\phi)$, where $\mathbf{q}_\phi^{\epsilon}$ is of order $\epsilon$. Each phase provides an anchored CV ($h_{\alpha}$, $h_{\beta}$, and $h_{\gamma}$ for AgI). A single anchored CV such as $h_{\alpha}$ measures departure from the $\alpha$ basin without specifying which competing polymorph is approached, so a biased trajectory can terminate in either competitor.
We bias $h_\alpha$ alone and identify the product of each run from the stacking of the iodine sublattice (SI Sec.~S3).

We also constructed CVs by replacing the trained GET encoder with the frozen encoder of MACE-MH-1, the potential driving the MD simulations.
In this setup, the embedding $\mathbf{h}_i$ of atom $i$ is the concatenation of its $\mathrm{SE}(3)$-invariant node features after the first and second message-passing layers of MACE-MH-1 (512 channels each, from which MACE-MH-1 also reads out the site energy).
Only the VAMPNet head, here with one hidden layer of width 64 (SI Sec.~S3\,B), is trained on these 1024-dimensional features, using the same loss, training trajectories, state count ($K_\mathrm{Ag}=14$), and lag time ($\tau = 300$~fs) as for the GET-based model.
During biased MD, the CV and its bias forces are evaluated in the same MACE-MH-1 forward pass that computes the potential energy and atomic forces.

All enhanced-sampling simulations of AgI and the unbiased trajectories used to train their joint EntroVAMP models were performed in LAMMPS in the $NPT$ ensemble at 380~K and 1~bar with a 1~fs time step using the MatPES r$^2$SCAN (\texttt{matpes\_r2scan}) head of MACE-MH-1 (SI Tables~S2 and S6). To generate the training data, unbiased trajectories of the $\alpha$, $\gamma$, and $\beta$ phases of AgI (512, 512, and 432 atoms, respectively) were run for 1.1~ns each, and the joint model ($K_\mathrm{Ag}=14$) was trained at a lag time of $\tau = 300$~fs. We performed well-tempered metadynamics simulations with a bias factor of 100 and Gaussian hills deposited every 25~fs, using PLUMED~\cite{tribelloPLUMED2014} for the CVs of the GET-based encoder and an in-house implementation of the same PLUMED actions for the CVs of the frozen MACE-MH-1 encoder (SI Sec.~S3\,B).
Using the GET-based encoder, we conducted 20 independent 1~ns trajectories biasing $h_\alpha$.
Using the frozen MACE-MH-1 encoder, we conducted 32 additional 1~ns trajectories biasing $h_\alpha$ with the same bias factor, Gaussian height, and deposition interval (Extended Data Fig.~\ref{edfig:agi_es_latent}). The product phases of all trajectories are summarized in SI Table~S7.

\section*{Data Availability}
Supporting data and source data for the figures, including the MD input files, initial structures and trained EntroVAMP models, are available at \url{https://github.com/AM3GroupHub/EntroVAMP}.

\section*{Code Availability}
The code base is open-sourced at \url{https://github.com/AM3GroupHub/EntroVAMP}.

\section*{Acknowledgments}

This work was supported in part by the AI2050 program at Schmidt Sciences (Grant G-25-69776 and Computing Fund), the NUS High Performance Computing facility (NUSREC-HPC-00001), and the National Supercomputing Center of Singapore (NSCC).

\section*{Author contributions}
P.Z. conceived and designed the research. 
M.C. developed the EntroVAMP model and codebase. X.W. and Z.Z. contributed to computational workflows and simulations, and interpreted the results.
M.C., X.W. and P.Z. wrote the manuscript with input from Z.Z. and Q.L.
P.Z. and Q.L. offered insight throughout the project. All authors contributed to discussions.

\section*{Competing interests}
The authors declare no competing interests.

\bibliography{references}

\clearpage
\setcounter{figure}{0}
\renewcommand{\theHfigure}{ED\arabic{figure}} 
\renewcommand{\figurename}{Extended Data Fig.}
\crefalias{figure}{edfigure}

\begin{figure*}[t]
\centering
\includegraphics[width=\textwidth]{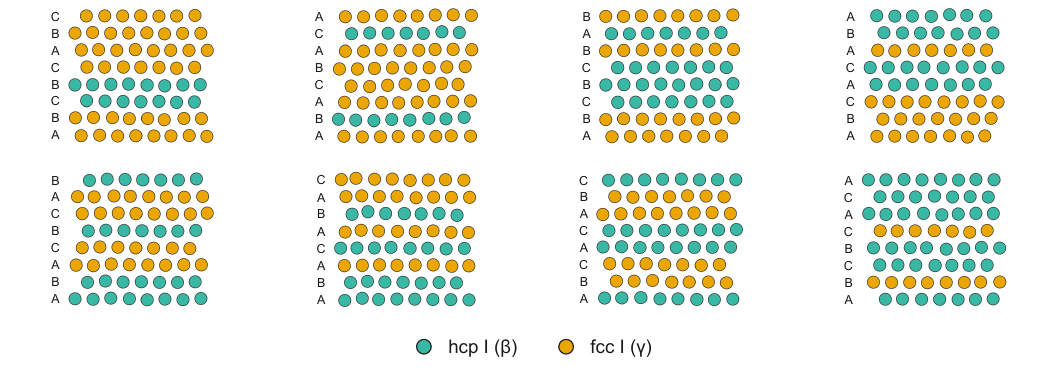}
\caption{\textbf{Representative stacking sequences of stacking-disordered AgI from enhanced sampling at 380~K with EntroVAMP.}
Snapshots show the iodine sublattice (hcp layers in teal, fcc layers in gold) for additional stacking-disordered products sampled during the metadynamics runs of Fig.~\ref{fig:agi_es}, viewed and labeled as in Fig.~\ref{fig:agi_es}c.\label{edfig:agi_sd_gallery_gnn}}
\end{figure*}

\begin{figure*}[t]
\centering
\includegraphics{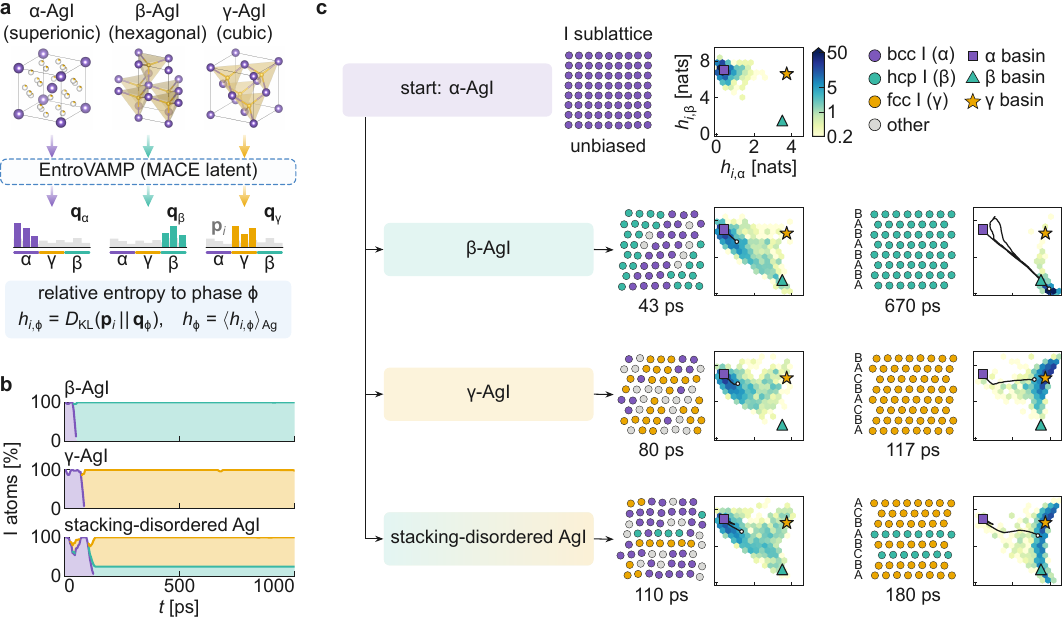}
\caption{\textbf{Enhanced sampling of phase transitions in AgI at 380~K with a frozen MACE-MH-1 encoder.}
Panels follow Fig.~\ref{fig:agi_es}, except that the trained GET encoder of EntroVAMP is replaced by the frozen encoder of MACE-MH-1, the potential driving the MD simulations.
The VAMPNet head takes as input the invariant node features that MACE-MH-1 passes to its site-energy readout.
All three runs bias $h_\alpha$, as in Fig.~\ref{fig:agi_es}.\label{edfig:agi_es_latent}}
\end{figure*}

\end{document}